Ernst Niederleithinger[1], Simon Gardner[2], Thomas Kind[1], Ralf Kaiser[2,3], Marcel Grunwald[1], Guangliang Yang[2,3], Bernhard Redmer[1], Anja Waske[1], Frank Mielentz[1], Ute Effner[1], Christian Köpp[1], Anthony Clarkson[2,3], Francis Thomson[2,3], Matthew Ryan[4]

# Muon tomography of a reinforced concrete block – first experimental proof of concept

[1] Bundesanstalt für Materialforschung und -prüfung (BAM), Berlin, Germany
[2] School of Physics & Astronomy, University of Glasgow, University Avenue, Glasgow, G12 8QQ, United Kingdom
[3] Lynkeos Technology Ltd., No 11 The Square, University of Glasgow, Glasgow, G12 8QQ, United Kingdom
[4] National Nuclear Laboratory, Central Laboratory, Sellafield, Seascale, Cumbria CA20 1PG, United Kingdom

## Abstract

Quality assurance and condition assessment of concrete structures is an important topic world-wide due to the ageing infrastructure and increasing traffic demands. Common topics include, but are not limited to, localisation of rebar or tendon ducts, geometrical irregularities, cracks, voids, honeycombing or other flaws. Non-destructive techniques such as ultrasound or radar have found regular, successful practical application but sometimes suffer from limited resolution and accuracy, imaging artefacts or restrictions in detecting certain features. Until the 1980s X-ray transmission was used in case of special demands and showed a resolution much higher than other NDT techniques. However, due to safety concerns and cost issues, this method is almost never used anymore.

Muon tomography has received much attention recently. Novel detectors for cosmic muons and tomographic imaging algorithms have opened up new fields of application, such as the investigation of freight containers for contraband or the assessment of the contents of radioactive waste containers. But Muon imaging also has the potential to fill some of the gaps currently existing in concrete NDT. As a first step towards practical use and as a proof of concept we used an existing system to image the interior of a reference reinforced 600 kg concrete block. Even with a yet not optimized setup for this kind of investigation, the muon imaging results show more resolution and less distortion compared to ultrasonic and radar imaging. The data acquisition takes more time and signals contain more noise, but the images allowed to detect the same important features that are visible in conventional high energy x-ray tomography.

In our experiment, we have shown the tremendous potential of muon imaging for concrete inspection. The next steps include the development of mobile detectors and optimising acquisition and imaging parameters.

## Funding

The work of the University of Glasgow has been supported by funding from STFC and EPSRC via the University of Glasgow Impact Accelerator Account.

## Conflicts of interest/Competing interests

The authors state no conflict of interests.

## Availability of data and material

The data are available from the authors on request. The reference concrete block is accessible at BAM.

## Code availability

No special code has been developed for this study

# 1 Introduction

The continuous availability of the European road transport network is one of the essential prerequisites for mobility and economic growth in the EU and world-wide. EU road infrastructure is getting older and suffers from aging issues with a large part of it already approaching the end of its life. According to the European Union Road Federation, the network had a length of 5.5 million km and a value of 8000 billion Euros in 2018, the latter declining [1]. In Germany, 10% of the bridges (bridge deck area considered) under federal administration were rated with a condition "less than sufficient" [2]. In France, about 25.000 bridges are prone to structural health issues which affect both safety and accessibility [3]. They are subject to serious fatigue problems, due to the increase of freight volumes (and traffic) with ever greater overall vehicle weights. Bridges and roads allow individual mobility and the supply of private households and the economy, but their aging and upcoming fatigue problems are leading to progressive degradation of bridge structures and thus to safety and reliability problems. These effects are further intensified by technological developments in heavy goods vehicle traffic (e.g. road trains, platooning etc.). Damage to structures that is usually only detected at a late stage can have far-reaching consequences for traffic. In the worst case, the total failure of a structure can lead to the complete inaccessibility of entire road sections in the traffic network, as illustrated by the collapse of the Polcevera Viaduct (aka Morandi Bridge) in Genoa in August 2018, leading to multiple deaths and resulting in a significant loss of gross domestic product. Traditional approaches for assessing the condition of transport infrastructure are based on structural inspections at fixed or adjustable time intervals. They are inadequate for an efficient inspection of the transport infrastructure assets, which after all amount to about 40% of the total European assets [1]. Only an efficient inspection, preferably permanently under flowing traffic will give infrastructure owners and managers the right picture to prioritize their maintenance operations.

There are already a number of Non-Destructive Testing (NDT) methods that provide engineers with tools to inspect aging infrastructure [4],[5],[6],[7],[8]. Standard technologies for structure assessments are ultrasonic methods and ground penetrating radar. For specific tests, e.g. reference measurements with very high resolution, X-ray radiography is also used. However, all of these techniques have their limitations. Ground penetrating radar is a very rapid and effective inspection method and is very sensitive for metal detection, but in concrete constructions the penetration depth and resolution of ground penetrating radar depend on the frequency of the radar used: Low frequencies can penetrate up to 1.5 m with resolutions of several cm and high frequencies can reach resolutions of several mm but the penetration depth is limited to about 40 cm. Ultrasonic echo instruments show greater penetration depths (around 1 m in commercial applications) but have a resolution of at best 1 cm. In addition, it is often not possible to inspect beyond the first reinforcement layers due to reflections. Ultrasonic echo methods are excellent to detect voids, cracks or delaminations, but cannot image features behind these obstacles. X-ray radiography (including variants such as X-ray tomography and laminography) can provide images with excellent resolution. Depending on the radiation energy, the penetrated thickness of a concrete structure can be up to 1 m with a spatial resolution of a few mm. However, when using X-ray radiography, attention must always be paid to compliance with the radiation protection regulations. In practice this often means that X-ray radiography cannot be applied.

Muon tomography, a purely passive technique using natural cosmic background radiation as a source, has the potential to overcome some of these issues [9]. Showers of high energy particles, including muons, are constantly created by collisions between cosmic rays and the upper atmosphere. The muons from these showers are highly penetrative and can pass through tens and hundreds of meters of rock before coming to rest and decaying. As cosmic muons are a naturally occurring radiation there are no costs or energy requirements in generating them, and because no additional radiation is generated there is no safety concern. It is a passive imaging system. There are two types of muon imaging techniques. The first is muon absorption imaging (or muon radiography), and the second is muon multiple scattering imaging (or muon tomography). While muon radiography uses one detector (or a set of detectors on the same plane) to detect muons after passing the object of interest, muon tomography uses two detectors (or two sets of detectors on different planes) to detect the muons before and after passing through the object of interest. The latter allows volumetric reconstruction of the object's scattering properties, resulting in high resolution 3D images [10].

The absorption of naturally occurring cosmic-ray muons was first used to investigate complex structures over 60 years ago by British physicist E. P. George, when he determined the weight of ice above a mining tunnel in Australia [11]. Imaging the interior of a volcano by muography was shown in 2001 [12]. Fifteen years ago, researchers at Los Alamos National Laboratory demonstrated that the Coulomb-scattering of the muon could be exploited to identify high-density, high-atomic number (Z) material within large, shielded transport containers [13], [14],Since this discovery, the non-destructive testing field of cosmic-ray muography has developed and, in recent years, has experienced an exponential growth with more than 40 research groups and projects active in over 20 countries throughout the world. In 2017 the topic received great attention with the publication of the high-profile measurements from within the great pyramid of Khufu in Egypt that indicated the presence of a previously unknown chamber [15]. In recent years, half a dozen companies have formed to commercialise muography imaging technology for a variety of different applications including nuclear contraband detection for national security, brownfield mineral exploration and nuclear waste characterisation. This industrial application is the subject of this current work, carried out by researchers at the University of Glasgow and its spin-out company Lynkeos Technology Ltd.

The idea of using muon tomography as a tool for the inspection of concrete structures including a concept for developing a suitable detecting system was submitted to the EC research program Horizon 2020 in early 2019 by the authors. The same idea was developed by another research group and successfully explored by simulations [16]. So far, all other similar applications reported have been limited to simulations or conceptual designs due to the lack of suitable detectors[17],[18].

As a proof of concept experiment, muon tomography of a reference reinforced concrete block produced by BAM was carried out at the University of Glasgow in September 2019 using a detector system originally developed to examine radioactive waste containers. To our knowledge, this is the first ever experiment of this kind. The results have been compared to several state-of-the-art techniques for concrete NDT at BAM. We have chosen the ultrasonic echo and radar echo techniques as the most used, state of the art methods for non-destructive investigation of the internal geometry of concrete structures. X-Ray laminography was performed as a reference.

The paper is organized as follows: In the next section ("Materials and Methods"), the reference concrete block is introduced as well as the various methods and devices used for examination, including techniques for data processing and imaging. In the "Results" section the images produced by the various techniques are shown and compared for three different horizontal cross sections of the reference block. In "Discussion" the advantages and limitations are compared. Conclusion and outlook section finalise this paper.

## 2  Materials and Methods

### 2.1  The reference concrete block "Radarplatte"

The reference concrete block "Radarplatte" (radar slab) was produced for training purposes. In its volume of 1.2 m x 1.2 m x 0.2 m four different targets were placed, which are typical for reinforced concrete structures (Fig. 1). Near to the top and the back, reinforcement bar mats were placed, each covering just about 50 % of the area and overlapping on about 25% of the area. Concrete cover is about 30 mm. The mat at the top has a mesh size of 150 mm and on the back of 100 mm. In between these two-reinforcement bar mats, an empty tendon duct with a diameter of 65 mm and a concrete cover of 90 mm was placed. Finally, a 50 mm thick Styrofoam block was inserted at the bottom to simulate a flaw at the backwall of the block.

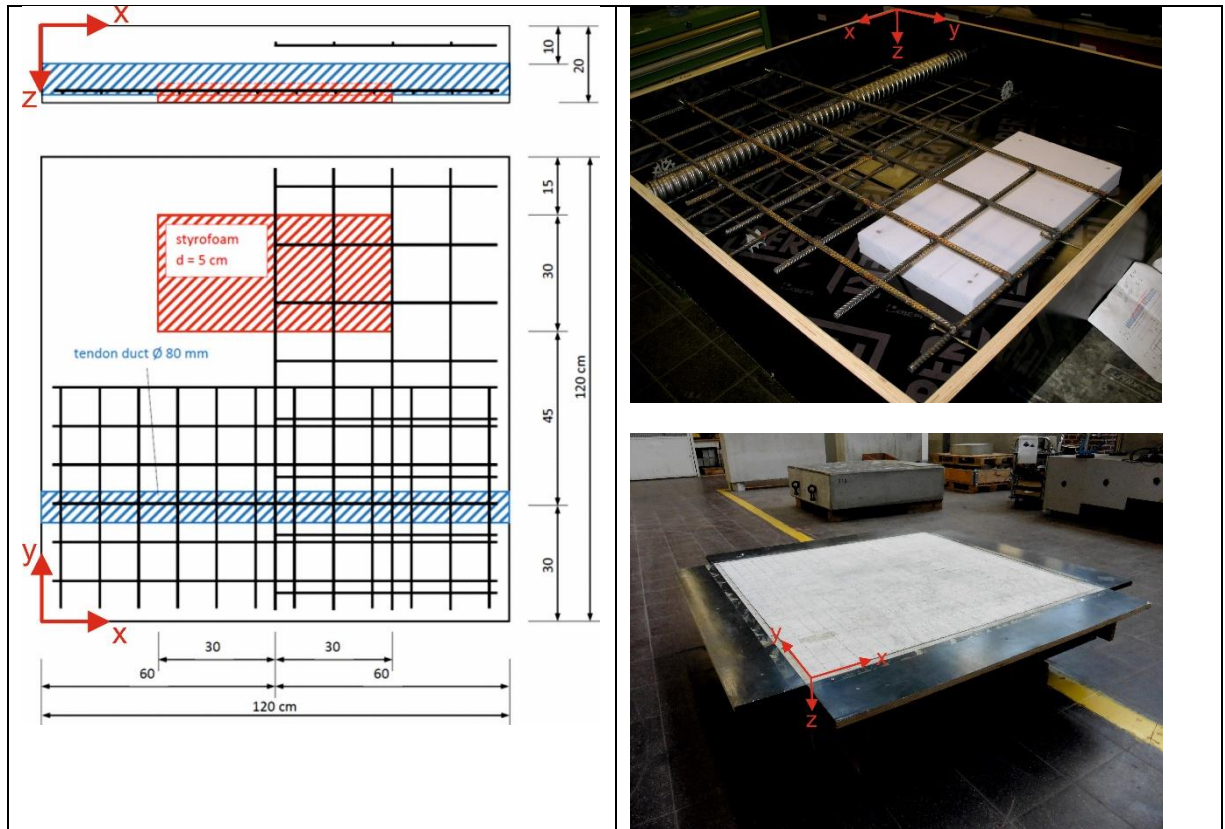

*Fig. 1 The reference concrete block "Radarplatte". Left: Design cross-section and view from above. Right: Pictures before and after concreting*

### 2.2 Muon tomography

Muon tomography is a technique, which is used to reconstruct 3D density maps of volumes using the Coulomb scattering of muons [9],[10]. By measuring the tracks of muons as they enter and exit the volume, an estimate of the average magnitude of scattering occurring in discrete volume elements can be calculated. Due to their high average energy of several GeV, i.e. 10,000 times that of typical X-rays, and due to the way muons interact with matter, they are highly penetrating and can pass through tens and hundreds of meters of rock (or concrete). The primary advantages of using muon tomography over other methods are this penetration depth and the fact that it is entirely passive while also being non-destructive. The comparatively long time it takes to make a measurement using cosmic-ray muons, can be considered its main detractor; millions of muons are required to create a high-resolution image and the flux of muons at sea level is around 170 Hz/m$^2$. This means that in practice data needs to be collected continuously for days or even weeks. The flux of muons also has a strong angular dependence, characterised by $\cos^2\theta$ to the vertical, which leads to a better imaging resolution in the horizontal plane than in the vertical direction.

Note, that the term tomography is used in differently in muon imaging and X-ray radiography related literature. In NDT standards, including those for X-ray imaging, tomography refers to imaging methods using 360° ray coverage, generated by rotating the object or the source/detector setup. The correct term for imaging planar objects with limited ray coverage (access just from two sides of the object) is laminography. However, to be consistent with the respective literature we are staying with the term tomography for the muon imaging method used in this research.

The Lynkeos Muon Imaging System (MIS) used for the investigation of the Radarplatte consists of 4 detector modules each containing 2 orthogonal layers of scintillating fibres from which a space point can be determined (Fig. 2, Table 1). Two modules placed above the volume are used to reconstruct the incident muon tracks and two below for the outgoing, scattered tracks.

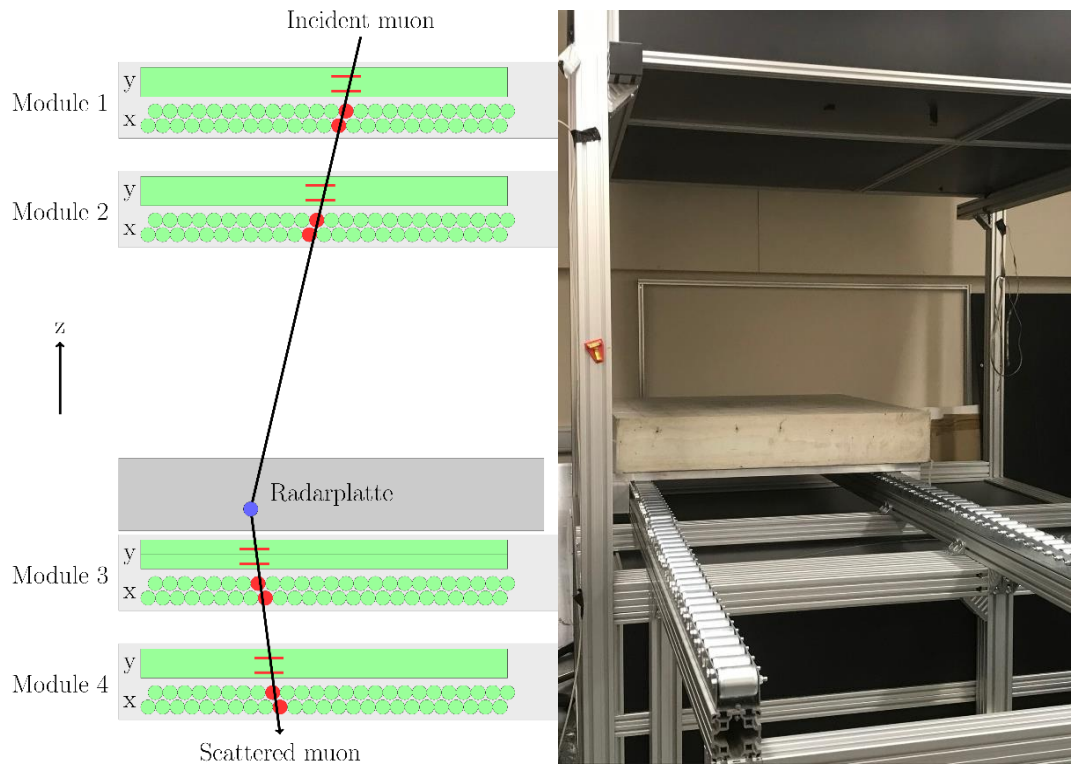

***Fig. 2*** *Muon Imaging. Left: Principle of muon tomography using two detectors each before and after the object to trace muon flight paths and scattering, Right: The "Radarplatte" test object inside the Lynkeos Muon Imaging System (MIS)*

The active area of the MIS modules is 1 m by 1 m allowing objects having smaller cross sections to be imaged. The horizontal resolution of the MIS is limited by the 2 mm diameter of the scintillating fibres used in the detectors; these are triangularly packed in two sublayers allowing an effective resolution of less than 2 mm where muons pass through neighbouring fibres. The vertical resolution of the reconstructed image is of the order of 4 cm due to the angular acceptance of the detector being limited to near vertical tracks.

| Source | Cosmic ray muons (1-100 GeV) |
|---|---|
| Detector | Lynkeos Muon Imaging System (MIS) |
| | 1024 x 1024 Fibres (Resolution <2 mm) |
| | Exposure time = 1203 hours |
| | Trigger rate = 11 Hz |
| Reconstruction | Voxel size 3.4 mm x 3.4 mm x 10 mm |
| | Size 300 x 300 Pixel and 178 slices (1060 x 1060 x 1780 mm³) |

*Table 1: Experimental parameters for muon tomography*

The volume between the top and bottom detectors is divided into voxels. The value attributed to each voxel is calculated based on the average measured scatter of the set of muons which pass through the voxel. The voxel value is expected to increase with the density of the volume it relates to.

In total 23 million muon tracks were used to reconstruct the tomographic image of the concrete sample. These muons were detected during the continuous running of the MIS for 1203 hours between 23[rd] September and 12[th] November 2019.

### 2.3   Radar

Radar is a non-destructive testing tool for the investigation of Civil Engineering structures and is based on the transmission and reception of electromagnetic waves [19],[20],[21],[22]. The received signal gives information about the internal structure of an investigated object by reflecting the transmitted electromagnetic wave back at objects which are conductive like metals or have different dielectric properties like concrete and air. The distance and position of objects can be derived from the received signal and material properties by analysing the size and shape of the received signal. The equipment for radar is designed for using different broadband antennas with a relative bandwidth of about 100%. The typical centre frequency of these antennas, which are applied for the investigation reinforced concrete buildings, lies in a range of 1 to 3 GHz. The penetration depth decreases with higher frequencies and the resolution increases simultaneously.

The radar data were collected by guiding a radar antenna manually along parallel lines with 5 cm distances, parallel to the sides of the Radarplatte (Fig. 3). A distance wheel encoder was connected to the antenna to collect A-scans every 2.5 mm along the line. Data were acquired both in x- and y-direction.

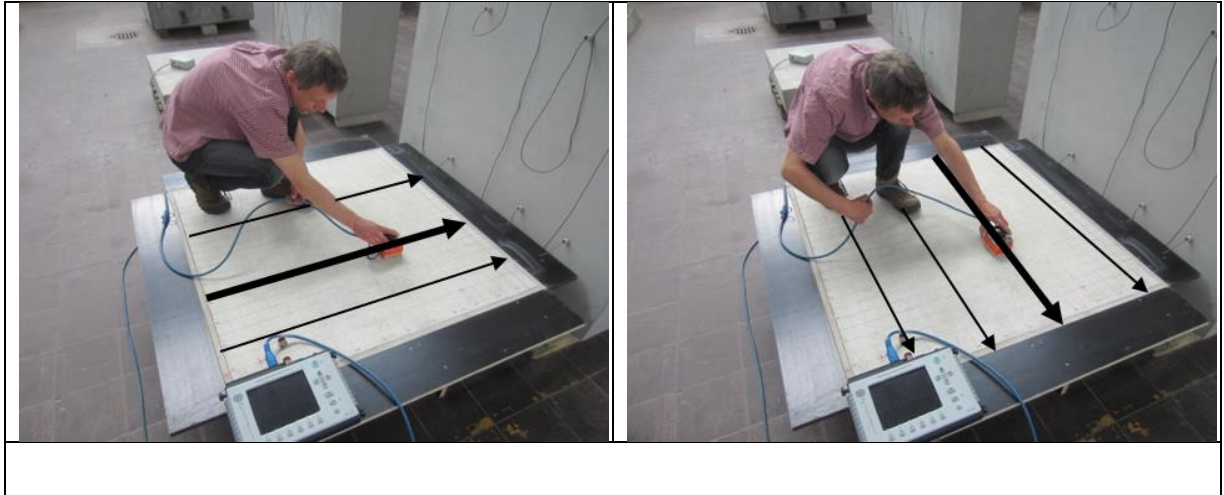

*Fig. 3 Radar data acquisition and the orientation of the two sets of measurement lines on top of the "Radarplatte".*

An antenna with a centre frequency of 2 GHz was used, which was connected to a radar control unit (GSSI SIR3000, Fig. 3). The collected data were analysed using the proprietary software of the manufacturer. The main processing steps included the application of a travel time dependent gain, 2D reconstruction by Kirchhoff migration and the calculation of the envelope function by Hilbert transform. All processed profiles (x- and y-direction) were assembled to a three-dimensional data cube. The travel time axis of the three-dimensional data cube was transformed to a depth axis by using a constant wave propagation speed. Three depth slices were generated for the depth 5 cm, 12 cm and 17 cm. Each depth slice was averaged over 1 cm in z-direction.

## 2.4   Ultrasound

Ultrasonic echo measurements have been established for the investigation of concrete constructions for about 25 years. Point contact shear wave transducers without the need of a coupling agent were introduced into practical application in the mid-1990s and are now almost exclusively used in the testing of concrete components. Today's commercial devices consist of two to sixteen arrays of three to twelve coupled transducers. The frequency range is in between 40 kHz and 60 kHz (3 dB attenuation), leading to a resolution in the centimeter-range. Reflections are recorded from elastic impedance contrasts within the object (e.g. concrete-steel, concrete-air) and from its boundaries. At interfaces to air, the energy is almost totally reflected, shadowing all features behind such interfaces. This means, that ultrasonic echo techniques can't image features behind e.g. delamination. Aggregates and larger pores are causing scattering of ultrasonic waves, leading to an inherent level of structural noise. Depth of penetration is limited to values around 1 m, depending on the degree of reinforcement, porosity, aggregate size and other factors. The method is mainly used for thickness measurement, geometry evaluation, detection of larger rebar, tendon ducts, voids, cracks and delaminations. The state of the art is described e.g. in [23],[24],[25].

The ultrasonic data were collected using an automated scanning system (Fig. 4) developed by BAM. A transducer array ("probe" in Fig. 4) with 12 shear wave point contact transducers each for transmitting and receiving was used. The probe was connected to a setup based on a custom mode pulse generator and commercial data acquisition equipment. Two data sets were collected on 2 cm by 2 cm grid using with different orientations of the probe (x- and y-polarization). The data sets were processed and imaged using the software InterSAFT developed by University of Kassel [26]. The most important parameters are shown in Table 2. Other than with radar, the resulting data volumes were kept separate. In the results section, the data set showing the greater response to the features in question is displayed.

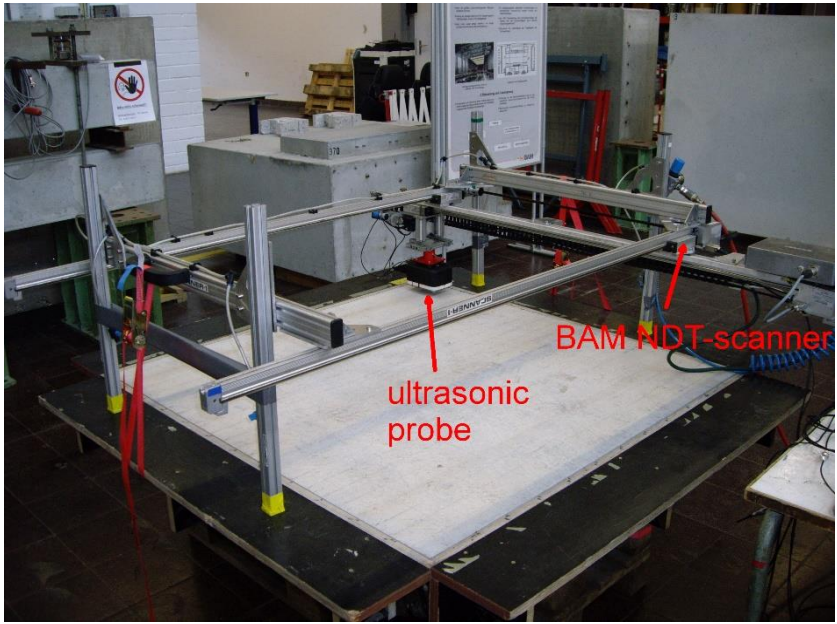

***Fig. 4*** *BAM NDT scanner with ultrasonic shear wave probe (Acsys M2503) mounted on the "Radarplatte".*

| Probe | Acsys M2503 |
|---|---|
| Data Acquisition | BAM proprietary setup using NI components, BAM NDT scanner |
| Pulse center frequency | 50 kHz |
| Sample rate | 1 MHz |
| Samples | 1000 |
| Point (A-scan) distance | 2 cm by 2 cm |
| Number of A-scans | 51 x 51 (10 cm offset to the edges) |
| Polarisation | x- and y-direction |
| Reconstruction | SAFT (Software InterSAFT) |

Table 2: Experimental parameters for Ultrasound

## 2.5 X-Ray Laminography

While X-ray computed tomography (CT) has been widely used as a common non-destructive imaging technique, it is not well-suited to visualise internal structures of large and flat objects. This is because CT requires a full rotation of the object and the object must fit in the field-of-view of the digital X-ray detector. Laminography or tomosynthesis are applied on laterally extended planar objects with large aspect ratios for example pipeline, large concrete sample and rotors of wind power plants, where the 360º image acquisition around the object is not physically possible.

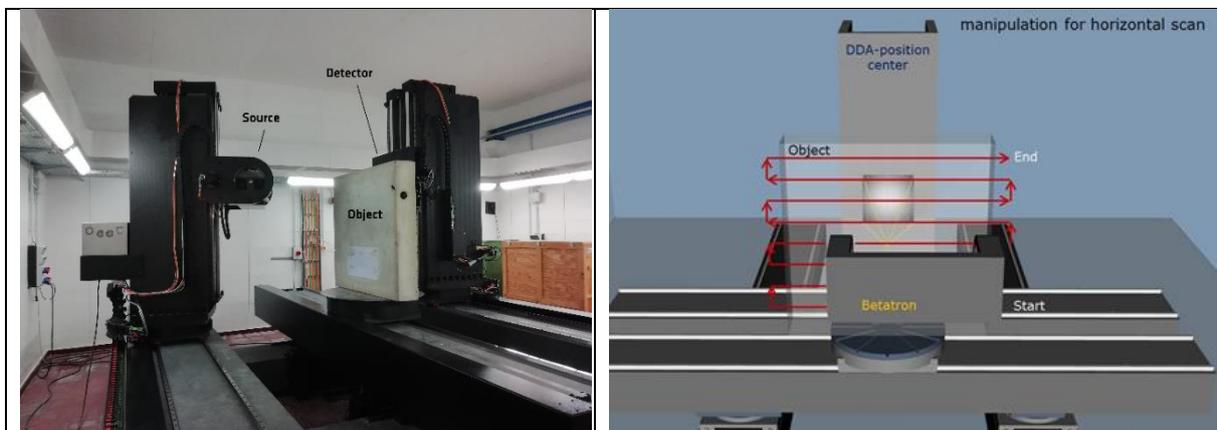

***Fig. 5*** *Left: HEXYTech-equipment for X-Ray Laminography of large and thick-walled test object. Right: Sketch of the horizontal scanning procedure. For each DDA position the source and detector move synchronously along the marked paths (red) relative to the object movement and several hundred projections are recorded for each (horizontal) path.*

In classical laminography, which is based on a relative motion of the X-ray source, the detector and the object can be set up in different geometrical arrangements. The X-ray source and the detector are either moved synchronously on circular paths around the object, a so-called rotational laminography, or are simply moved in opposite directions in the case of translational laminography.

Planar tomography (PT) as a special case of coplanar translation laminography was used to investigate the concrete plates using the HEXYTech equipment of BAM (Fig. 5). The radiation source (X-ray tube, gamma source, accelerator) and detector (e.g. matrix detector) are moved synchronised and parallel to the object, whereas the object remains stationary and is irradiated by X-rays at various positions (usually several hundred) and the radiation that penetrates the object is recorded by a detector. The digitally stored projections contain 3-dimensional information of the object. A 3D volume data set of the studied object is reconstructed from the projections by a filtered backpropagation algorithm (FBP) that is adapted to the specific geometry of the laminography arrangements. In the volume data set of the concrete plate, different features can be detected, e.g. reinforcement, concrete matrix, cracks and air inclusions [27]. Unfortunately, X-ray data is not free of artefacts that result from irregular illumination or the high pass filter used by the reconstruction algorithm to enhance the edges in the projection data. These artefacts complicate the quantitative analysis of the cracks.

The experimental parameters of the planar tomography of the Radarplatte are described in Table 3. The pixel resolution of the digital detector was 400 µm. The total number of single projections was approx. 7300.

| Source | Betatron JME 7.5 MeV , Focus Spot Size = 0.3 x 3 mm |
|---|---|
| Detector | Perkin Elmer XRD |
| | 2048 x 2048 Pixel (Resolution 200 µm) |
| | 1024 x 1024 Pixel (Resolution 400 µm) |
| | Exposure time = 1000 ms |
| | Number of single frames = 6 |
| Reconstruction | Voxel size 0.5 mm x 0.5 mm x 1.5 mm |
| | Size 2460 x 1170 Pixel and 200 slices (1230 x 585 x 300 mm³) |

Table 3: Experimental parameters for X-ray planar tomography

## 3  Experimental results

The experimental results have been acquired and processed as described in the previous section, resulting in 3D voxel datasets of the investigated volume. The volumes have been geometrically referenced to the upper main surface of the object (z = 0 m). The x and y axis are along the larger boundaries of the objects. To evaluate and compare the results of the investigations, three depth sections parallel to the upper surface have been extracted at depths of 5 cm, 12 cm and 17 cm below the upper surface.

Note, that these depth sections are not necessarily produced by the respective values at the precise depth but may be averaged over a certain depth interval. This is a usual procedure for ultrasonic and radar images acquired on concrete to smooth the structural noise caused the inherent inhomogeneous nature of concrete. Some of the datasets are limited to certain parts of the object due to experimental limitations of the devices used. X-ray tomography misses 10 cm of the upper part of the block. The muon tomography is currently limited to 1 m by 1 m and misses a small margin left and top, and a larger one bottom and right. In addition, the object was inserted into the muon detection system with an offset of the edges of 0.6 degrees, which was not corrected in the imaging process. These limitations can be overcome by optimising the setup.

The uppermost of the three depth levels discussed here intersects with the upper reinforcement mesh (Fig. 6 a). Muon tomography (Fig. 6 b) shows all rebars clearly, proving the sub-cm resolution of this technique. Some of the features in larger depth show up slightly (tendon duct and Styrofoam plate) as bright shadows, which is typical for experiments of the transmission tomography type. The boundaries show dark colours, most probably reconstruction artefacts. Radar (Fig. 6 c) shows all rebars clearly, but the signatures are wider than the actual bars. Ultrasound (Fig. 6 d) is not able to image the reinforcement in this case as it is beyond the resolution limit for the setup used here. X-ray laminography (Fig. 6 e) shows the clearest picture of all technologies and gives a correct estimate for the rebar diameter (12 mm). Shadows from deeper features and boundary effects are similar to the ones seen in muon tomography. Interestingly, the reinforcement grid is distorted in all successful technologies (e.

g. non-constant distance between bars). These distortions were verified by detailed photographs of the grid before concreting.

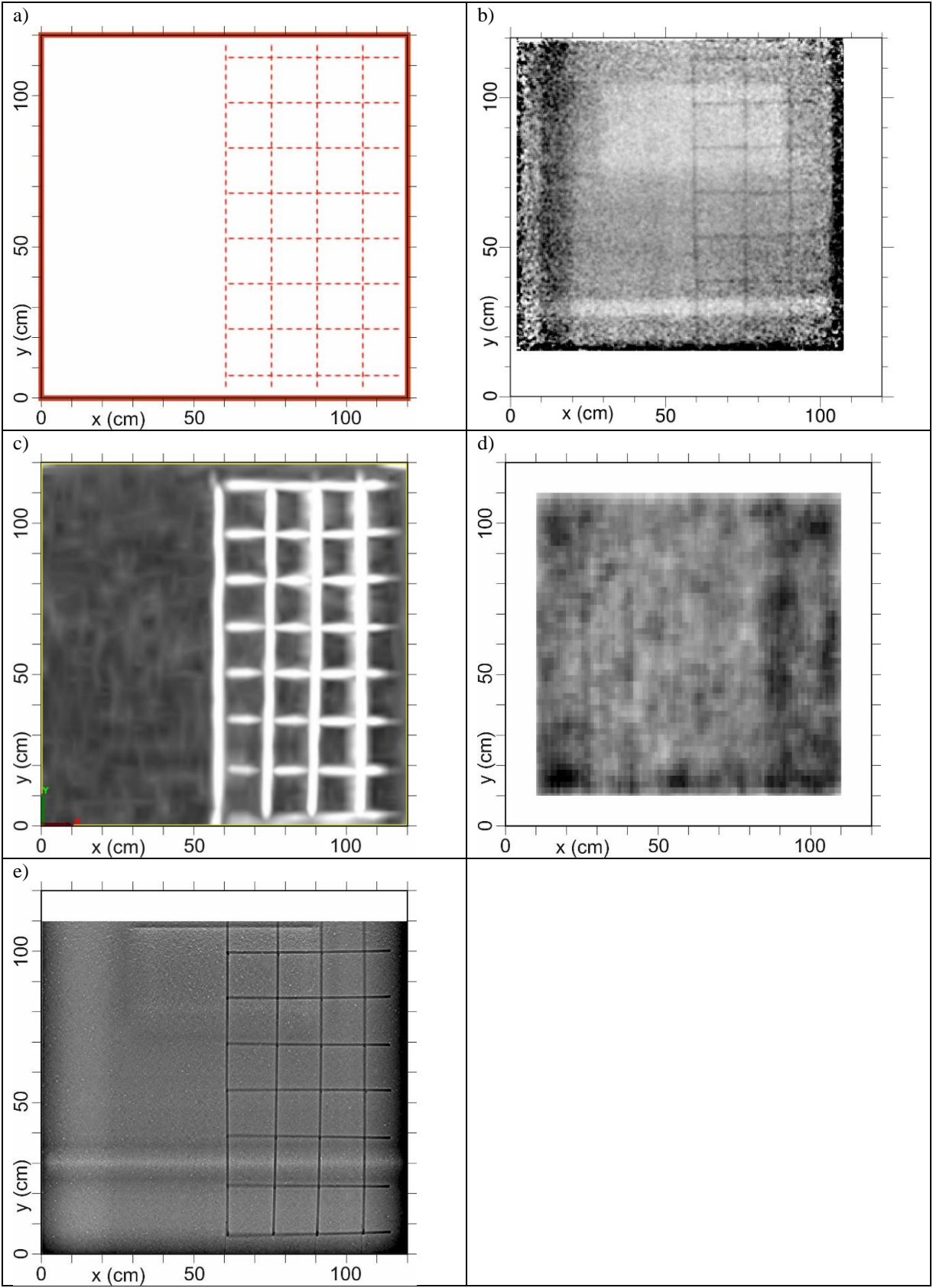

*Fig. 6 The horizontal cross-sections, depth 5 cm (upper reinforcement). a) design with upper reinforcement, b) muon tomography, c) radar, d) ultrasound (y-Polarization) and e) X-ray laminography*

The second depth level (12 cm) intersects with the center of the tendon duct (Fig. 7 a). All technologies are able to image this feature, alas, with different clarity and level of detail. While X-Ray laminography (Fig. 7 e) shows even the undulations of the corrugated pipe, all other methods images are more or less straight shaped and show a significant level of noise. Radar (Fig. 7 d) shows artefacts from the reinforcement layer above, which is typical for echo techniques. Muon tomography shows the tendon duct with a clarity comparable to radar and ultrasound. As in X-ray laminography, artefacts from features above and below are present.

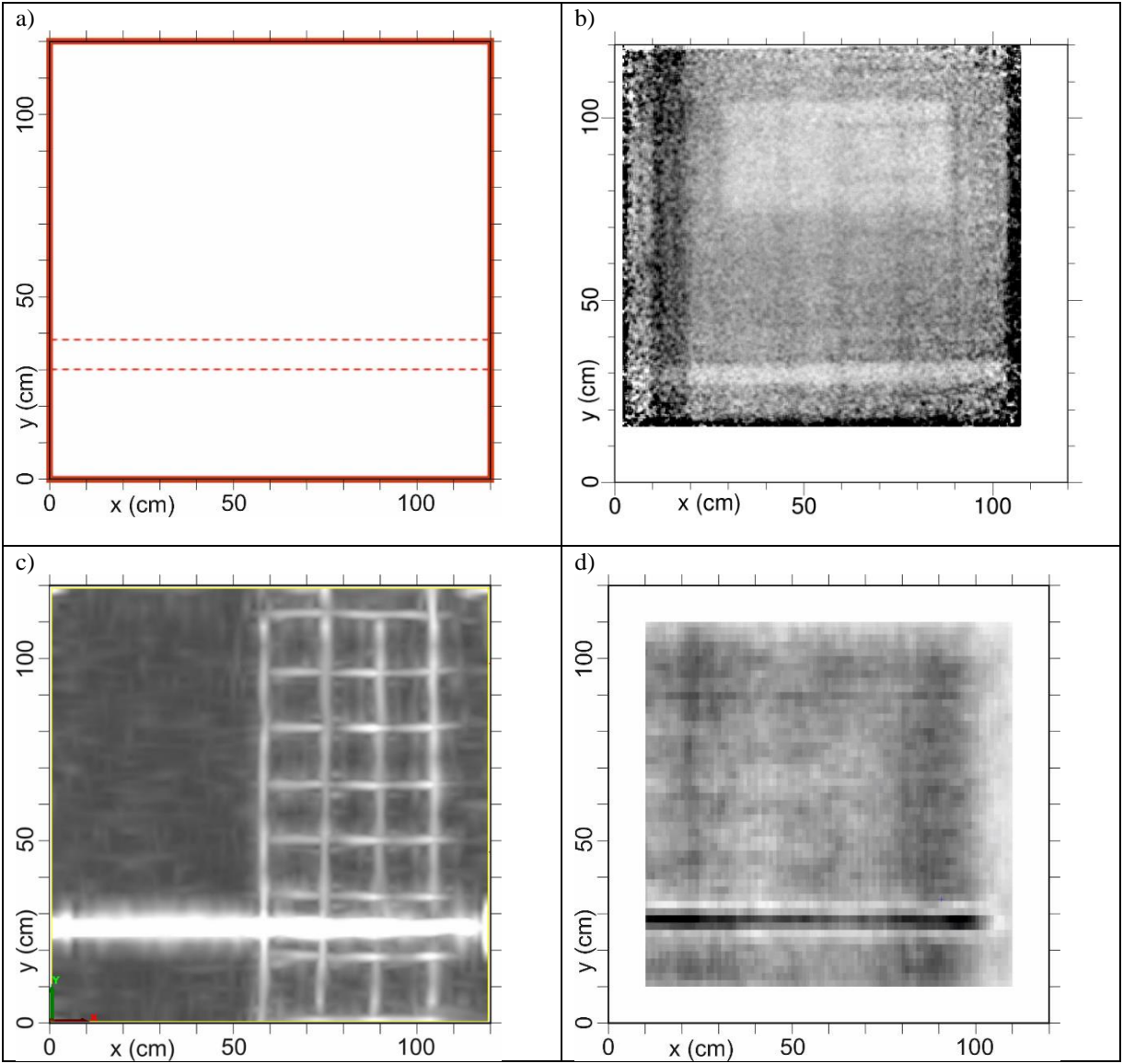

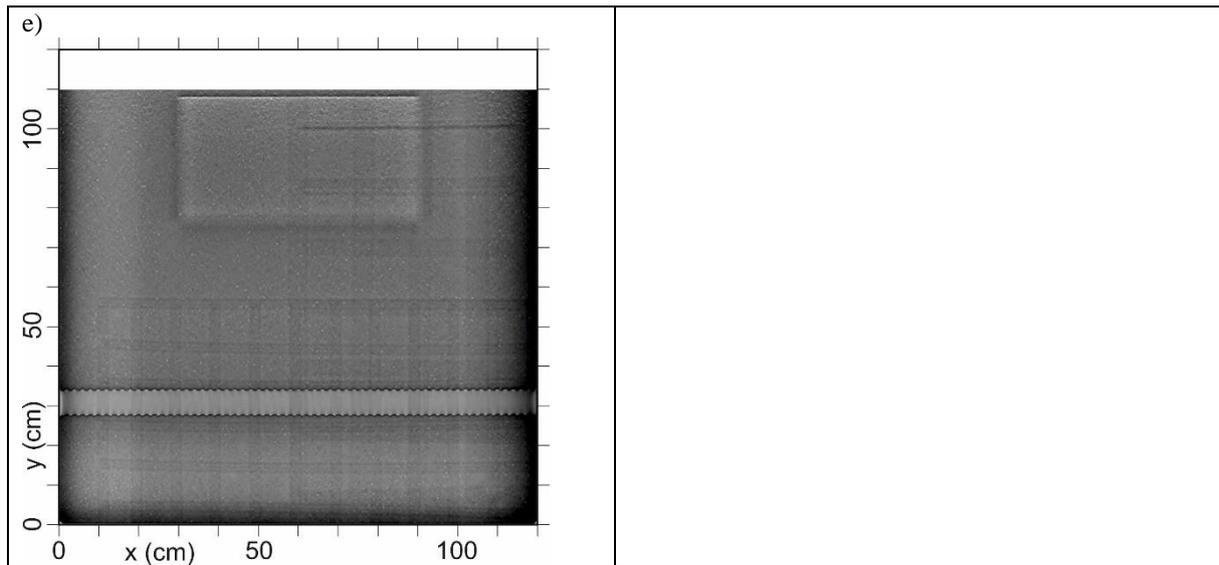

*Fig. 7 The horizontal cross-sections, depth 12 cm (tendon duct). a) design with upper reinforcement, b) muon tomography, c) radar, d) ultrasound (x-polarization) and e) X-ray laminography*

The third depth level studied intersects with the lower reinforcement and the Styrofoam block (Fig. 8 a). The Styrofoam block is imaged by all techniques, while the reinforcement is missed by ultrasound (Fig. 8 d) as for the upper reinforcement. The radar image (Fig. 8 c) is partially distorted by artefacts caused by the features above this depth level. These features show up in the muon tomography (Fig. b) and X-ray laminography (Fig. 8 e) images as well, but with lesser effect on the image of the objects, which are actually at this depth level.

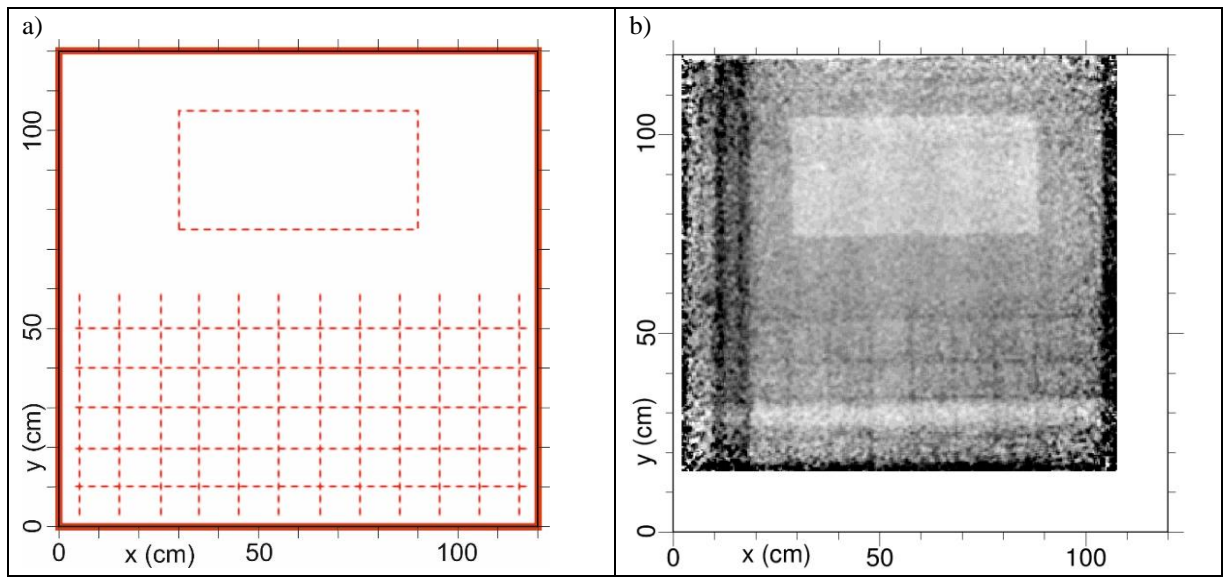

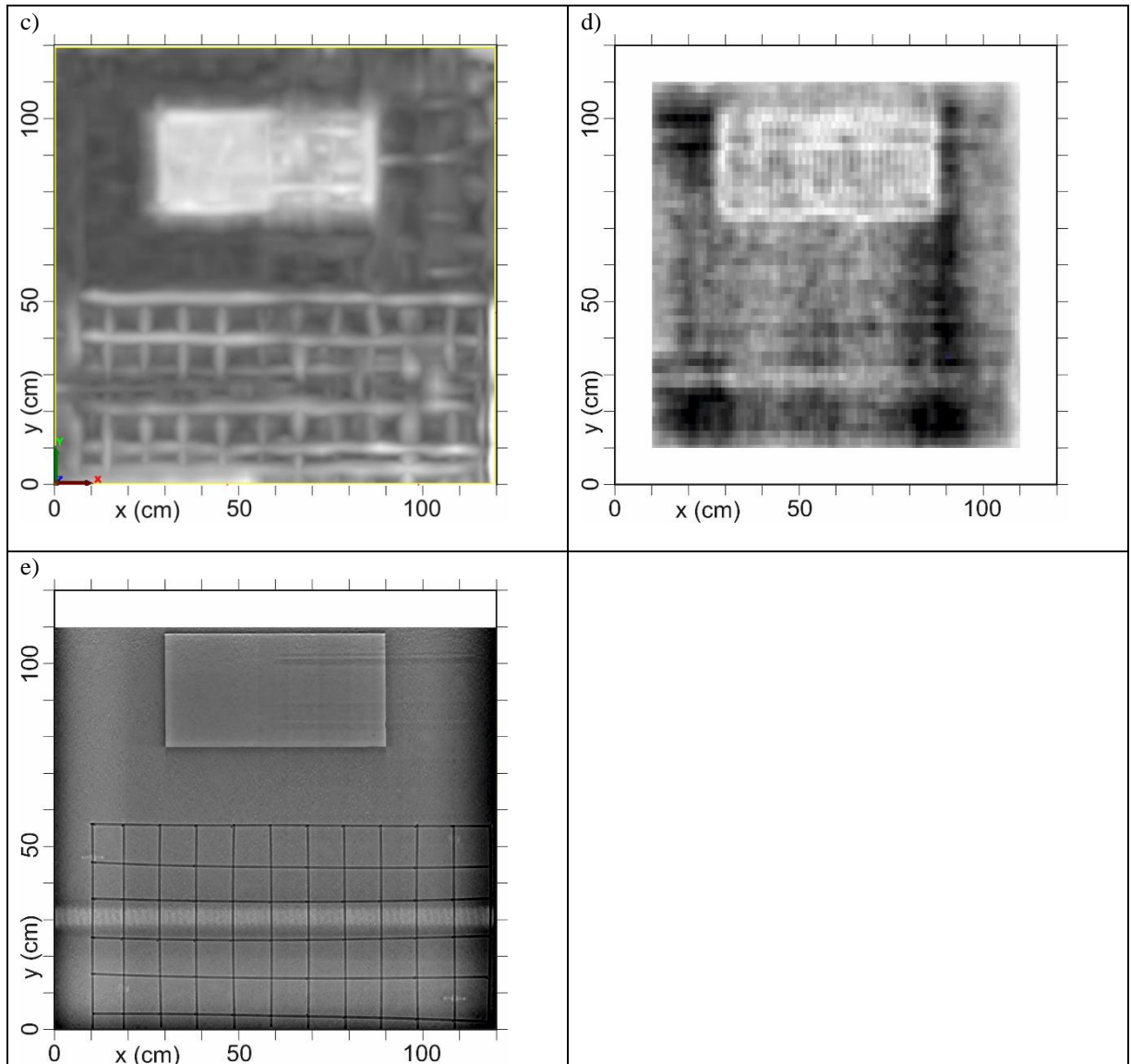

*Fig. 8* The horizontal cross-sections, depth 17 cm (lower reinforcement and Styrofoam plate). a) design with upper reinforcement, b) muon tomography, c) radar, d) ultrasound (y-Polarization) and e) X-ray laminography

## 4 Discussion

The muon tomography images, acquired with a setup which is not yet optimized for concrete inspection, have shown that high density and low density features in concrete objects can be imaged at the correct position by this emerging technology and be distinguished without further data processing (low density/air: bright, high density/steel: dark). A characterization of objects would be possible with radar and ultrasound as well, but this would involve additional non-standard processing steps. The resolution of muon tomography exceeds the one of ultrasound and radar, at least for the scenario investigated here. Some artefacts are present in the images. Some (e.g. the shadows of objects above and below the depth level under consideration) are due to two inherent limitations of the technology: First, tomographic reconstruction algorithms may produce artefacts and are "smearing" anomalies in case of limited angular coverage. Second, the angle of the incident muons varies just between -30° and +30° from the vertical axis due to the geometrical acceptance of the detector system. This is limiting the vertical resolution.

In addition to all the features present in the concrete slab, an additional high-density vertical band is present towards the left edge of all the muon tomography images. This is the shadow of part of the support structure which held the concrete sample in position during the measurement. The image edges are noisier and more sensitive to misalignment due to the limited acceptance of events only from vertical muons.

The comparison of muon tomography images with those of X-Ray laminography show similarities (shadowing effects from above and below, edge effects, different signature of low- and high-density objects). The resolution of the images as well the low noise level are distinct advantages of active X-Ray technologies. Still, the quality of the muon tomography images is beyond our expectations for a first-ever experiment on reinforced concrete using a new technology in a non-optimized setup.

## 5   Conclusion and Outlook

Our first-ever experiment with muon tomography of a reinforced concrete block was successful. All built-in features were detected and correctly identified. The positioning accuracy matches the one of radar and ultrasound while the resolution seems to be even better. Note that the data shown here required a recording time in the order of weeks while the acquisition of ultrasonic data was performed within about two hours and of the radar data in less than 30 minutes. Equipment costs and requirements regarding operator skills are currently much lower for radar and ultrasound as well. Muon tomography at its current state does not reach the resolution and noise level of X-Ray tomography. However, it does not require any radiation safety measures on site.

We are optimistic that muon tomography can be developed into a useful tool for non-destructive structural investigations to fill the gaps in technologies currently used on site. The highest priority to progress this exiting technology is the development of an efficient and affordable mobile muon detector. Measurement and processing parameters still have to be optimised. A thorough validation of the technology and its possibilities and limitations has to follow. We foresee that a combination of different technologies using methods from data fusion and/or improvements of the reconstruction software using machine leaning will be important steps on the way to becoming a standard technology.

## 6   Acknowledgements

The work of the University of Glasgow has been supported by funding from STFC and EPSRC via the University of Glasgow Impact Accelerator Account.

## 7   References

[1] European Union Road Federation: Road asset Management – An ERF position paper for maintening and improving and efficient road network. http://erf.be/wp-content/uploads/2018/07/Road-Asset-Management-for-web-site.pdf, downloaded 2020-07-10.

[2] German Road Research Institute (BASt), 2020. Bridge Statistics.https://www.bast.de/BASt_2017/DE/Statistik/Bruecken/Brueckenstatistik.pdf , downloaded 2020-07-10

[3] Rapport d'information de MM. Patrick CHAIZE et Michel DAGBERT, fait au nom de la commission de l'aménagement du territoire et du développement durable n° 609 (2018-2019) - 26 juin 2019

[4] Breysse, Denys, und International Union of Laboratories and Experts in Construction Materials, Systems and Structures, Hrsg. Non-Destructive Assessment of Concrete Structures: Reliability and Limits of Single and Combined Techniques ; State-of the Art Report of the RILEM Technical Committee 207-INR. RILEM State-of-the-Art Reports 1. Dordrecht: Springer, 2012.

[5] Helmerich, Rosemarie, Ernst Niederleithinger, Daniel Algernon, Doreen Streicher, und Herbert Wiggenhauser. „Bridge Inspection and Condition Assessment in Europe". *Transportation Research Record: Journal of the Transportation Research Board* 2044, Nr. 1 (Dezember 2008): 31–38. https://doi.org/10.3141/2044-04.

[6] Maierhofer, Christiane, Hans-Wolf Reinhardt, und Gerd Dobmann, Hrsg. Non-Destructive Evaluation of Reinforced Concrete Structures. Vol. 1: Deterioration Processes and Standard Test Methods. Boca Raton; Oxford: CRC Press ; Woodhead, 2010.

[7] Buyukozturk, O., Hrsg. Nondestructive testing of materials and structures: proceedings of NDTMS-11, Istanbul, Turkey, May 15-18, 2011. RILEM bookseries, v. 6. Dordrecht ; London: Springer, 2013.

[8] Balayssac, Jean-Paul, und Vincent Garnier, Hrsg. *Non-destructive testing and evaluation of civil engineering structures*. Structures durability in civil engineering set. London, UK : Kidlington, Oxford, UK: ISTE Press ; Elsevier, 2018.

[9] Kaiser, Ralf. „Muography: Overview and Future Directions". *Philosophical Transactions of the Royal Society A: Mathematical, Physical and Engineering Sciences* 377, Nr. 2137 (28. Januar 2019): 20180049. https://doi.org/10.1098/rsta.2018.0049.

[10] Yang, Guangliang, Tony Clarkson, Simon Gardner, David Ireland, Ralf Kaiser, David Mahon, Ramsey Al Jebali, Craig Shearer, und Matthew Ryan. „Novel Muon Imaging Techniques". *Philosophical*

*Transactions of the Royal Society A: Mathematical, Physical and Engineering Sciences* 377, Nr. 2137 (28. Januar 2019): 20180062. https://doi.org/10.1098/rsta.2018.0062.

[11] George EP. 1955 Cosmic rays measure overburden of tunnel. *Commonwealth Engineer*, July 1, 455–457.

[12] Tanaka, H., K. Nagamine, N. Kawamura, S. N. Nakamura, K. Ishida, und K. Shimomura. „Development of the cosmic-ray muon detection system for probing internal-structure of a volcano". *Hyperfine Interactions* 138, Nr. 1/4 (2001): 521–26. https://doi.org/10.1023/A:1020843100008.

[13] Mahon, David, Anthony Clarkson, Simon Gardner, David Ireland, Ramsey Jebali, Ralf Kaiser, Matthew Ryan, Craig Shearer, und Guangliang Yang. „First-of-a-Kind Muography for Nuclear Waste Characterization". *Philosophical Transactions of the Royal Society A: Mathematical, Physical and Engineering Sciences* 377, Nr. 2137 (28. Januar 2019): 20180048. https://doi.org/10.1098/rsta.2018.0048.

[14] Simpson, Allan, Anthony Clarkson, Simon Gardner, Ramsey Al Jebali, Ralf Kaiser, David Mahon, Julian Roe, Matthew Ryan, Craig Shearer, und Guangliang Yang. „Muon Tomography for the Analysis of In-Container Vitrified Products". *Applied Radiation and Isotopes* 157 (März 2020): 109033. https://doi.org/10.1016/j.apradiso.2019.109033.

[15] Morishima, Kunihiro, Mitsuaki Kuno, Akira Nishio, Nobuko Kitagawa, Yuta Manabe, Masaki Moto, Fumihiko Takasaki, u. a. „Discovery of a Big Void in Khufu's Pyramid by Observation of Cosmic-Ray Muons". *Nature* 552, Nr. 7685 (Dezember 2017): 386–90. https://doi.org/10.1038/nature24647.

[16] Dobrowolska, Magdalena, Jaap Velthuis, Anna Kopp, Marcus Perry, und Philip Pearson. „Towards an application of muon scattering tomography as a technique for detecting rebars in concrete". *Smart Materials and Structures* 29, Nr. 5 (1. Mai 2020): 055015. https://doi.org/10.1088/1361-665X/ab7a3f.

[17] Zenoni A, et al. 2014. Historical building stability monitoring by means of a cosmic ray tracking system. In Proc. of the 4th Int. Conf. on Advancements in Nuclear Instrument Measurement Methods and their Applications. ANIMMA 2015, 20–24 April 2015 Lisbon: IEEE. (https://arxiv.org/pdf/1403.1709.pdf)

[18] Vanini, S., P. Calvini, P. Checchia, A. Rigoni Garola, J. Klinger, G. Zumerle, G. Bonomi, A. Donzella, und A. Zenoni. „Muography of Different Structures Using Muon Scattering and Absorption Algorithms". *Philosophical Transactions of the Royal Society A: Mathematical, Physical and Engineering Sciences* 377, Nr. 2137 (28. Januar 2019): 20180051. https://doi.org/10.1098/rsta.2018.0051.

[19] Daniels, D.J. Ground Penetrating Radar, ISBN 9780863413605, Institution of Engineering and Technology 2004

[20] Jol, Harry M.Ground Penetrating Radar Theory and Applications, ISBN, 0080951848, Elsevier, 2008

[21] J. Hugenschmidt, A. Kalogeropoulos, F. Soldovieri, G. Prisco (2010) Processing Strategies for high-resolution GPR Concrete Inspections, NDT & E International, Volume 43, Issue 4: 334-342

[22] Lai, Wallace Wai-Lok ; Dérobert, Xavier; Annan, Peter; A review of Ground Penetrating Radar application in civil engineering: A 30-year journey from Locating and Testing to Imaging and Diagnosis, NDT & E International, Volume 96, 2018, Pages 58-78, ISSN 0963-8695

[23] Krause, Martin, Klaus Mayer, Martin Friese, Boris Milmann, Frank Mielentz, und Gregor Ballier. „Progress in Ultrasonic Tendon Duct Imaging". *European Journal of Environmental and Civil Engineering* 15, Nr. 4 (Januar 2011): 461–85. https://doi.org/10.1080/19648189.2011.9693341.

[24] Wiggenhauser, H.; Samokrutov, A. A.; Mayer, K.; Krause, M.; Alekhin, S.; Elkin, V.: LAUS – Large Aperture Ultrasonic System for Testing Thick Concrete Structures. In: Int. Symp. Non-Destructive Testing in Civil Engineering (NDT-CE), Berlin, 15 – 17.09.2015. Berlin: Bundesanstalt für Materialforschung und –prüfung (BAM), 2015, pp. 743 – 746 https://doi.org/10.1061/(ASCE)IS.1943-555X.0000314

**[25]** Krause, M.: Ultrasonic Echo – Physical principals and theory. In: Breysse, D. (ed.): Non-Destructive Assessment of Concrete Structures: Reliability and Limits of Single and Combined Techniques. State-of-the-Art Report of the RILEM Technical Committee 207-INR. Springer, 2012, Ch. 2.2, pp. 27 – 39. **DOI 10.1007/978-94-007-2736-6**

[26] Mayer, Klaus, und P.M. Cinta. „Mayer, K., Chinta, P.M.: User Guide of Graphical User Interface inter_saft". University of Kassel, Department of Computational Electronics and Photonics, 2012.

[27] Moosavi, R., Grunwald, M., Redmer, B. Crack detection in reinforced concrete. NDT & E, International, 109 (2020). https://doi.org/10.1016/j.ndteint.2019.102190